\documentclass[ reprint,
amsmath,amssymb,
 aps,
pra,
floatfix
]{revtex4-1}

\usepackage{graphicx}
\usepackage{dcolumn}
\usepackage{bm}
\usepackage{color}
\usepackage{soul}

\begin{document}


\title{Pulsed versus continuous wave operation of a ring Stark decelerator}

\author{Yomay Shyur}
 \email{yomay.shyur@colorado.edu}
\author{Jason A. Bossert}
\author{H. J. Lewandowski}
\affiliation{JILA, NIST, and Department of Physics, University of Colorado, Boulder, Colorado 80309-0440, USA}%

\date{\today}

\begin{abstract}
Stark deceleration is a technique that uses time-varying inhomogeneous electric fields to  decelerate polar molecules for various molecular beam and trapping experiments. New ring-geometry Stark decelerators with continuously varying voltages offer a method to produce a more intense source of molecules in a technique called traveling-wave Stark deceleration. However, this type of deceleration is more experimentally challenging than the more typically used crossed-pin geometry decelerators with pulsed voltages. Here, we present an experimental realization of a ring-geometry Stark decelerator using either continuously varying or discrete voltages. Pulsed-ring Stark deceleration using discrete voltages is easier to implement and, under certain circumstances, is more efficient than traveling-wave Stark deceleration. A comparison of experimental and simulated results between traveling-wave and pulsed-ring Stark deceleration is presented along with a simple model for determining when each mode is more efficient. 
\end{abstract}

\maketitle


\section{\label{sec:intro}Introduction}

Stark deceleration uses spatially inhomogeneous and time-varying electric fields to decelerate neutral polar molecules, and produces a beam of slow molecules with a tunable average final velocity. These beams can be used for high-resolution spectroscopy \cite{vanVeldhoven2004}, collision physics \cite{Kirste2012}, measuring radiative lifetimes of molecules \cite{vdM2005,Gilijamse2007}, and tests of fundamental physics \cite{Hudson2006,Bethlem2008,Tarbutt2009}. The molecules from a Stark decelerator can also be loaded into  electrostatic \cite{Bethlem2000,vanVeldhoven2006,Gilijamse2010}, magnetic \cite{Sawyer2007,Riedel2011}, and ac traps \cite{vanVeldhoven2005,Bethlem2006,Schnell2007}, or used in crossed beam  \cite{Gilijamse2006,vdM2006b,Scharfenberg2011,Scharfenberg2011b}, beam-trap \cite{Sawyer2008,Sawyer2011}, and co-trapped experiments \cite{Parazzoli2009}. Such experiments are often limited by the number of successfully decelerated molecules. Improvements to the decelerator efficiency, the percentage of molecules in the initial pulse that are successfully decelerated, would benefit these types of studies. 

The first Stark decelerator, referred to in this paper as the pulsed-pin Stark decelerator (PPSD), was constructed in 1999 by Bethlem \textit{et al.} \cite{Bethlem1999}. It used high-voltage switches to alternate between two static voltage configurations on pairs of crossed pin electrodes. Switching between these voltage configurations produces a discretely moving Stark potential well to decelerate molecules. Molecules only within a range of positions and velocities will be decelerated. This portion of phase space is called the phase-space acceptance. Ideally, it would be well filled with molecules. PPSD has well-characterized instabilities and loss mechanisms that are enhanced at low final velocities \cite{vdM2006}. Particularly, coupling between the longitudinal and transverse motion leads to unstable regions within phase-space acceptance. The decrease in phase-space acceptance drastically reduces the efficiency when decelerating to low velocities required for trapping, or when spending long times at velocities $<$100 m/s in the decelerator. Although different timing schemes and operating modes can improve the performance of the traditional Stark decelerator \cite{Scharfenberg2009,Parazzoli2009,Hou2013}, the crossed-pin electrode geometry fundamentally limits the efficiency due to longitudinal and transverse coupling.

The loss present in pin decelerators can be mitigated by using ring-shaped electrodes and a fundamentally different mode of applying voltages to the electrodes: chirped sinusoidally varying voltages. This produces a traveling-wave potential well that continuously guides or decelerates the molecular beam, as opposed the pulsed nature of PPSD. Traveling-wave Stark deceleration (TWSD) has two characteristic differences compared to PPSD: the cylindrically symmetric electrode geometry and  continuously, as opposed to discretely, varying voltages, which produce a genuine moving potential well. This leads to an inherently stable deceleration process that has a true three-dimensional (3D) trapping potential and a well-filled phase-space acceptance. These factors allow for more efficient deceleration due to the continuous and concentrically symmetric nature of this deceleration mode. Meek \textit{et al.} \cite{Meek2008} first demonstrated the idea for spatially modulated potential wells on a microstructured chip used to guide and then decelerate  metastable CO molecules \cite{Meek2009a,Meek2009b}. The first use of macroscopic traveling-wave Stark potentials with a ring-geometry decelerator was experimentally shown in 2010 by Osterwalder \textit{et al.} \cite{Osterwalder2010} using a beam of CO, which was decelerated from 288 m/s to 144 m/s.

While TWSD has many advantages, it is more challenging to implement than PPSD because it requires high-voltage analog amplifiers with demanding specifications to generate the moving Stark potential wells. The sinusoidal voltage must have an initial frequency such that the Stark potential well containing the molecules moves with the same initial velocity as the molecular beam. The frequency of the sinusoidal voltages then must be smoothly chirped down to a frequency such that the velocity of the Stark potential well formed by the voltages matches the desired final beam velocity. Thus, the amplifiers require a large bandwidth to decelerate a typical molecular beam to rest. The chirp rate of the sinusoidal voltage will be determined by the initial and final velocities and the length of the decelerator. For a fixed ring geometry, the magnitude of the maximum allowable deceleration is set by the bandwidth of the high-voltage amplifiers, maximum output voltage of the high-voltage amplifiers, and length of the decelerator.

Early TWSD experiments were limited by the lack of availability of high-voltage amplifiers with a large bandwidth. The amplifiers used to generate the voltages for the electrodes in the first TWSD experiment operated at $\pm$ 8-10 kV with 10-30 kHz sinusoidal voltages \cite{Osterwalder2010,Meek2011}. The limited bandwidth of the amplifiers meant that the decelerator could achieve only moderate deceleration and the molecules could not be brought to rest. Commercial amplifiers (Trek 5/80), which are capable of a larger frequency range (0-60 kHz) and outputs up to $\pm$5 kV, make it possible to decelerate a supersonic beam down to rest with a decelerator that is several meters long. For example, a beam of SrF seeded in Xe with an initial velocity of 290 m/s would require a $\sim$5 m long decelerator \cite{vandenberg2014}. Amplifiers with slightly limited maximum voltages, such as these, are particularly useful for the TWSD of heavy molecules used in precision measurements, since the electric fields produced are low enough to keep the molecules in the weak-field seeking regime, where the Stark energy increases with increasing electric field \cite{vandenberg2012}. To date, YbF \cite{Bulleid2012} and SrF \cite{vandenberg2014} have been decelerated using TWSD and the longest demonstrated decelerator (4 m) slowed SrF from 290 to 120 m/s using $\pm$5 kV amplifiers \cite{Mathavan2016}. 

To overcome the limitations of PPSD and the demanding amplifier specification of TWSD, a combination of both Stark deceleration methods can be used. In this combination technique, a pulsed-pin Stark decelerator decelerates molecules to longitudinal velocities around 100 m/s, and then a traveling-wave Stark decelerator decelerates the molecules to rest \cite{Quintero2013}. This scheme is able to take advantage of the cylindrical symmetry of the ring geometry at low velocities to avoid the low velocity losses of the pulsed-pin Stark decelerator, but it relaxes the requirement to have either extremely high maximum voltages or a very long decelerator to reach electrostatically trappable conditions. Using this combination technique, some have trapped NH$_3$/ND$_3$  and CH$_3$F inside the ring electrodes and have performed adiabatic cooling and trap manipulations inside the rings \cite{Quintero2013,Jansen2013,Quinter2014,Meng2015}. 

Complete TWSD of a supersonic beam moving at 300-450 m/s to rest has been challenging to achieve due to electronic and hardware requirements. An alternative running mode for operating a ring-geometry decelerator has been proposed by Hou \textit{et al.} \cite{Hou2016}, and uses pulsed voltages on ring electrodes. The duration of the pulses is varied in an identical manner to that of PPSD. This pulsed-ring Stark deceleration (PRSD) method takes advantage of the cylindrical symmetry provided by the rings and is simpler to implement than TWSD, since it uses commercial high-voltage switches instead of high-voltage analog amplifiers. In this paper, we present experimental results of this new PRSD scheme in comparison with TWSD, examine conditions where it is advantageous to run PRSD and TWSD, and discuss a simple method for determining which mode is preferable depending on the desired final molecule packet properties.

\section{\label{sec:expSetUp}Experiment}

\begin{figure}
\includegraphics[width=3.25in]{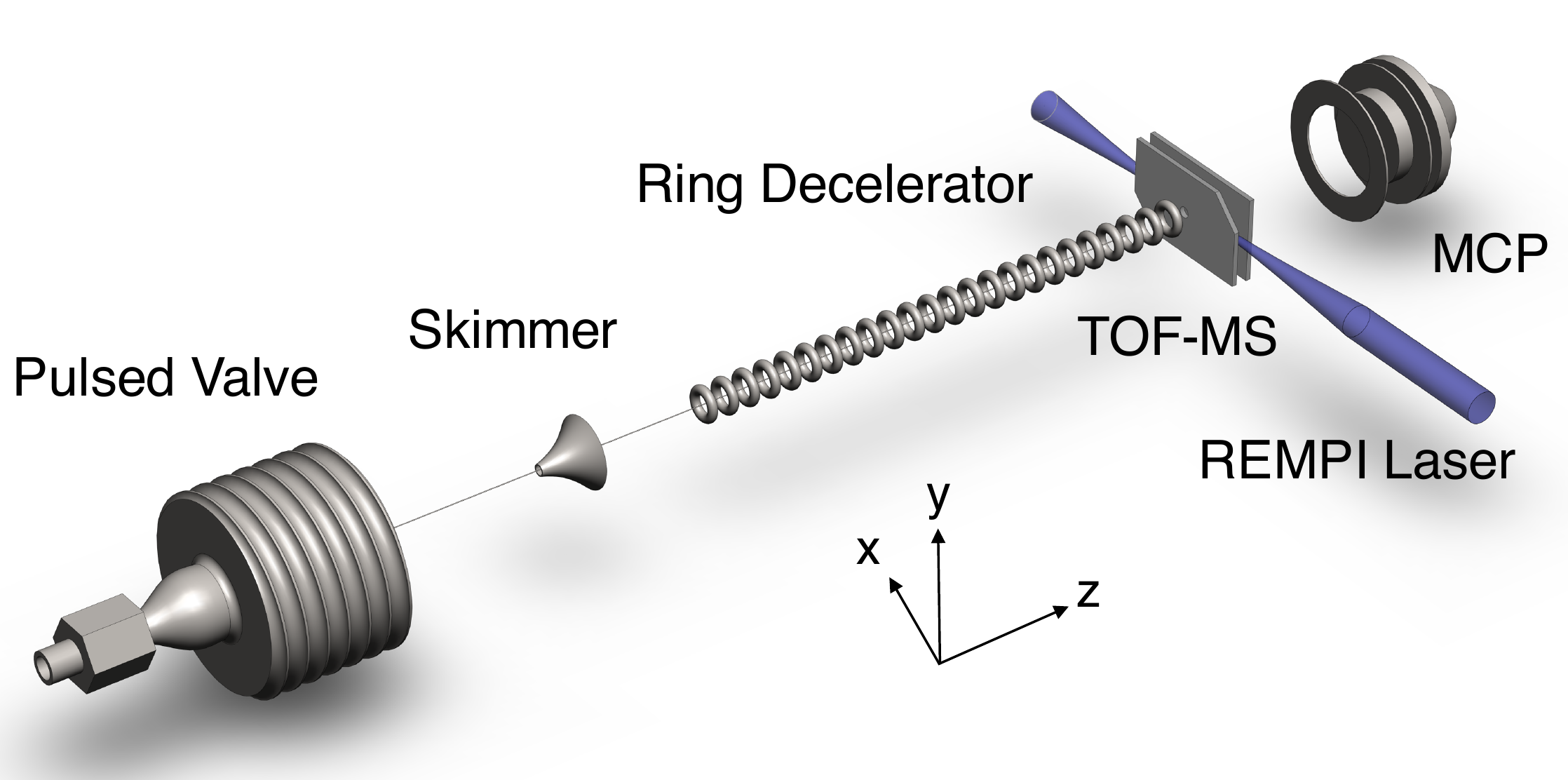}
\caption{\label{fig:expsetup} A schematic of the experimental setup which includes, from left to right, a pulsed valve, skimmer, ring decelerator, time-of-flight mass spectrometer (TOF-MS), detection laser, and microchannel plate detector (MCP). Not all decelerator rings are depicted. The full decelerator contains 624 rings. The longitudinal stainless steel mounting rods are also not shown. The entire system is contained inside a differentially pumped vacuum chamber.}
\end{figure}

The experimental setup for demonstrating both deceleration modes is shown in Fig. \ref{fig:expsetup}. Both experiments begin with a piezoelectric-actuated pulsed valve that creates a pulsed supersonic beam of 2$\%$ ND$_3$ in krypton. A skimmer $\sim$20 cm downstream of the valve collimates the molecular beam and allows for differential pumping between the source and decelerator chambers. The decelerator begins 3 mm behind the skimmer, and consists of 624 rings mounted in eight longitudinal stainless steel rods such that every eighth ring is electrically connected (similar to \cite{Osterwalder2010}). Each ring is made of 1.02 mm  diameter tantalum wire, and have a 4 mm inner diameter and center-to-center spacing of $l$, which for the decelerator discussed here is $l=$ 2.03 mm. The full decelerator is just over 1.25 meters long. For the work presented here, fully deuterated ammonia (ND$_3$) molecules are decelerated from 415 m/s down to final velocities of 220 m/s for TWSD and 150 m/s for PRSD. ND$_3$ molecules at the end of the decelerator are ionized in a 2+1 REMPI scheme using 317 nm photons \cite{Ashfold1998}. Two stainless steel plates, which are electrically isolated from the decelerator, are mounted after the rings. They form the time-of-flight mass spectrometer and accelerate ionized molecules into the microchannel plate detector, where the measured current is proportional to the number of ionized molecules. 

\begin{figure}
\includegraphics[width=3.25in]{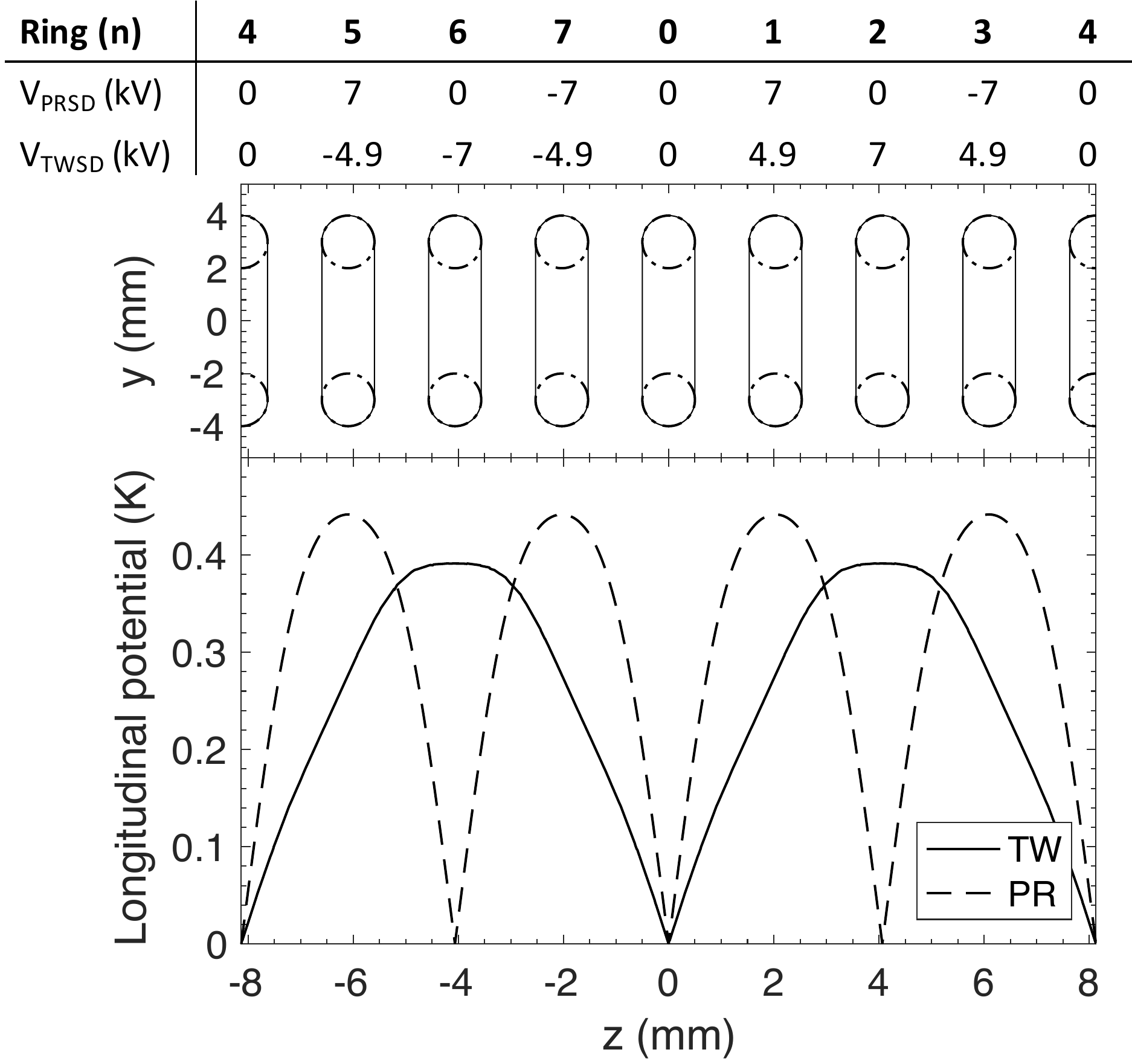}
\caption{\label{fig:potentials} Ring number, applied voltages, positions of electrodes, and longitudinal Stark potentials of ND$_3$ in units of Kelvin. The table and  top panel show the ring number $n$ and instantaneous voltages for both deceleration modes plotted below and the longitudinal cross section of the decelerator rings. The bottom panel shows the TWSD and PRSD Stark potentials along the center axis of the ring decelerator. Both Stark potentials are shown at a time when the bottom of the potential well is centered at a ring. }
\end{figure}

The only difference between the two deceleration modes, PRSD and TWSD, is the voltages that are applied to the electrodes. In the work presented here, both modes of deceleration use maximum voltages of $\pm$7 kV on the electrodes. PRSD uses four commercial high-voltage switches and TWSD uses eight custom home-made high-voltage amplifiers. Figure \ref{fig:potentials} shows a longitudinal cross section of the ring decelerator in the upper panel and the longitudinal Stark potential for each deceleration mode in the lower panel. The rings are labeled with their rod number, $n$, from zero through seven. The instantaneous voltages that create the given longitudinal Stark potentials are listed at the top. The left most and right most rings, $n=4$, are electrically connected, and the voltage and Stark potential pattern is repeated down the length of the decelerator. 

\subsection{\label{subsec:voltages}Deceleration Modes}
For the PRSD configuration, each rod is connected to one of four Behlke HTS 201-03-GSM-HFB high-voltage switches. These switches are used to produce two voltage configurations that alternate in time. The two voltage configurations produce two Stark potential well configurations, one of which is depicted in the bottom panel of Fig. \ref{fig:potentials}.  A detailed description of the electrode and voltage configuration can be found in \cite{Hou2016}. The time interval between switching from one configuration to another is dictated by the velocity of the molecular beam. These voltages produce an on-axis longitudinal Stark potential with the bottom of the well shifted by $l$ between the two configurations, similar to the pin decelerator. The timing sequence for switching the voltages on the electrodes is calculated in a similar fashion, using phase angles, as that of a pulsed-pin Stark decelerator \cite{Bethlem2002}. 

For the TWSD configuration, eight analog high-voltage amplifiers are used, one for each rod. These supply the rings with chirped sine wave voltages, which have a fixed phase offset relative to one another. The amplifiers (gain~=~12,000) are driven by an arbitrary waveform generator (GaGe CompuGen 8152). The high-voltage linear amplifiers used in this work were designed and constructed in-house. They were tested up to $\pm$10 kV and in the frequency range of 1-30 kHz on the bench, but were operated at $\pm$7kV  and in the frequency range of 13-25.5 kHz for this work. These specifications correspond to a molecular beam with velocities from 415 m/s down to 210 m/s. While the amplifiers have a bandwidth of 30 kHz down to DC, the full bandwidth was not utilized since the lower maximum voltage on the rings meant a shallower molecule trap. The set length of our decelerator and this shallower trap did not allow for deceleration down to rest for this particular experiment. However, with $\pm$10kV outputs from the amplifiers, the ND$_3$ molecules with an initial velocity of 415 m/s could be brought to rest. 

Each amplifier applies the chirped sine wave voltage $V_n$, for a given rod $n$, of
\begin{equation}
\label{eq:ampchirp}
V_n(t)=V_a \sin\left(\frac{2\pi n}{8}-\frac{2\pi}{8l}\left(v_it-\frac{1}{2} a t^2\right)\right),
\end{equation}
where $V_a$ is the maximum output voltage, $v_i$ is the initial velocity of the molecular beam, $a$ is the acceleration applied by the decelerator, and the $\frac{2\pi n}{8}$ term is the phase shift between rods. The acceleration is given by
\begin{equation}
\label{eq:accel}
a=\frac{v_f^2-v_i^2}{2L},
\end{equation}
where $v_f$ is the final velocity of the molecular beam and $L$ is the overall length of the decelerator. It is important to note that one period of the sine wave extends over eight ring electrodes and corresponds to two Stark potential wells (see Fig. \ref{fig:potentials}). This results from the Stark potential energy depending on the square of the magnitude of the electric field and not the sign. 

\section{\label{sec:accel}Simple model of deceleration}
\begin{figure}
\includegraphics[width=3.25in]{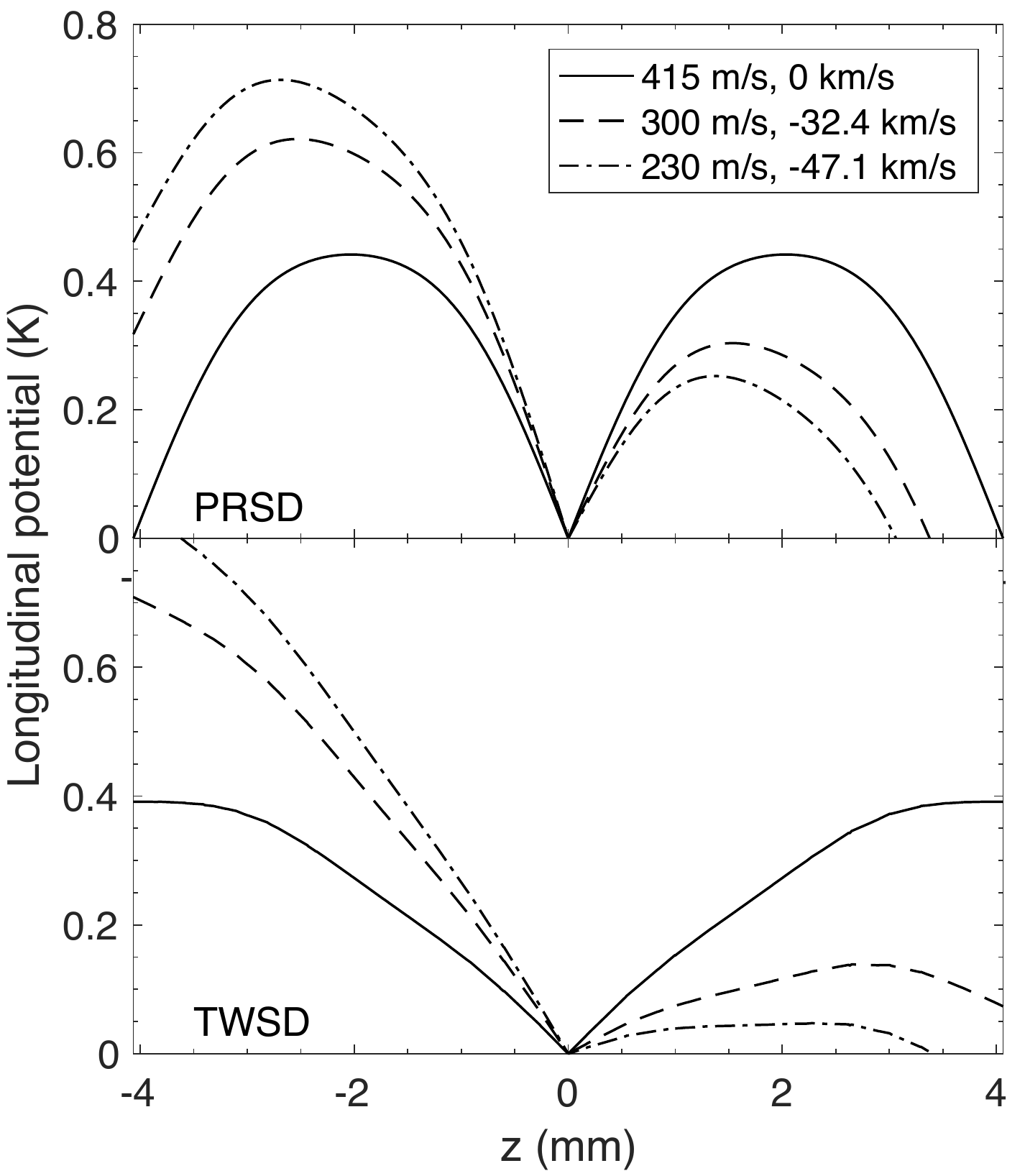}
\caption{\label{fig:tilt} The PRSD (top) and TWSD (bottom) Stark potential wells with applied linear fictitious potentials for $a=0$ km/s$^2$ (solid), $a=-32.5$ km/s$^2$ (dashed), and $a=-47.1$ km/s$^2$ (dot dashed). For a beam with $v_i=$415 m/s, this corresponds to $v_f=$300 m/s and $v_f=$230 m/s respectively. }
\end{figure}

To predict which molecules will be successfully decelerated (i.e. within the phase-space acceptance), we consider the effects of the acceleration on the Stark potential well. Deceleration of the Stark potential well is the equivalent of adding a fictitious potential $U_f=maz$ to the Stark potential in Fig. \ref{fig:potentials}, where $m$ is the mass of the molecule and $z$ is the longitudinal distance from the center of the Stark potential well. For both modes, the average acceleration is determined only by decelerator length, $v_i$, and $v_f$. The acceleration is applied constantly in TWSD, but is time-averaged in PRSD, where instantaneous accelerations are higher. Fig. \ref{fig:tilt} shows the central TWSD and PRSD Stark potential wells from Fig. \ref{fig:potentials} with an acceleration of 0 km/s$^2$, -32.5 km/s$^2$, and -47.1 km/s$^2$ added to the Stark potential. For a beam with $v_i=$415 m/s, this corresponds to $v_f=$415 m/s, $v_f=$300 m/s and $v_f=$230 m/s respectively. This additional fictitious potential lowers the downstream potential well wall since $a$ is negative, and the effect is referred to as the potential well tilt. (The tilt model of a Stark decelerator is discussed more in depth in \cite{Friedrich2004}.)  As $a$ increases, the well tips over more and cannot contain and decelerate as many molecules. An acceleration of -47.1 km/s$^2$ shows how the TWSD Stark potential well is aggressively tilted to the point of almost having no trap depth. Therefore, too much acceleration, and thus too much tilt, leads to a small phase-space acceptance and deceleration efficiency.

For either mode, too large of an acceleration means no molecules are phase stable inside the Stark potential well. The maximum acceleration where the phase-space acceptance is finite is set by the gradient and height of the Stark potential well. A larger Stark potential gradient allows for more acceleration before the well can no longer hold molecules. In Fig. \ref{fig:tilt} for $a=$-47.1 km/s$^2$, PRSD still has significant well depth, while TWSD has almost none. This is because PRSD has a larger Stark potential gradient. The Stark potential well gradient is set by the deceleration mode and physical decelerator geometry. Decelerating to low final velocities requires either a low initial velocity, which is set by the gas used in the supersonic expansion, or a long decelerator. For the same $V_a$ on the electrodes, the pulsed configuration produces a Stark potential well with a larger gradient and a smaller volume because the well only spans $2l$ while the traveling-wave well spans $4l$. Thus, the PRSD well is able to tilt more than the TWSD well even though it has a smaller volume.

\section{\label{sec:sims}Simulations} 

In addition to experimental results, we use Monte Carlo simulations to help interpret and illustrate the differences between the two operating modes. These semi-classical molecular trajectory simulations allow us to explore a wide parameter space that is not yet accessible in the current experimental setup. The simulation starts by creating a packet of molecules at the location of the pulsed valve. The molecules have a Gaussian distribution of initial positions and velocities. The molecular packet is allowed to propagate through a region with no electric field to the entrance of the decelerator. Once the molecules enter the decelerator, their trajectories are determined using a 3D matrix of accelerations derived from Stark potentials modeled by the commercial finite element solver COMSOL. Once the packet of molecules reaches the end of the decelerator  both the phase-space distribution (observation at a set time) and the time-of-flight information (observation at a set $z$-position) are recorded. 

Typical initial packets contain between 1 and 8 million molecules that are sampled from a distribution that has full-width half maximum of $\Delta x$~=~$\Delta y$~=~3 mm, $\Delta z$~=~30 mm, $\Delta V_x$~=~$\Delta V_y$~=~5 m/s, and $\Delta V_z$~=~30 m/s. This packet extends beyond the phase-space acceptance of the decelerator. 

Since the Stark potential well shapes for both deceleration modes are spatially periodic, the electric field models for determining the molecule accelerations inside the decelerator do not have to span the entire length of the decelerator. Instead, a modular approach can be taken. The position dependent molecular accelerations are calculated from electric field models for a unit cell of 3 (9) rings for PRSD (TWSD). Then, a shift of the longitudinal coordinates within the unit cell allows for molecules to propagate along the decelerator without a model of the entire decelerator. The number of position dependent 3D acceleration models required depends on the operating mode. For PRSD, there are only two voltage configurations and the shape of the Stark potential well is the same for both configurations, but shifted by $l$. Therefore, only one acceleration model is required and is longitudinally shifted by $l$ each time the high voltage is switched. For TWSD, there is a continuum of voltage configurations corresponding to the sinusoidally varying voltages on the rods. This is modeled using 25 time steps between the peak of the sine wave being on rod $n$ to the peak being on the subsequent rod $n+1$. Once the simulations have cycled through all 25 modeled time steps the entire model is longitudinally shifted by $l$. 

\section{\label{sec:exp}Experimental Results and Analysis}

\subsection{\label{subsec:7kV}Deceleration at 7kV}
\begin{figure*}
\includegraphics[width=\textwidth]{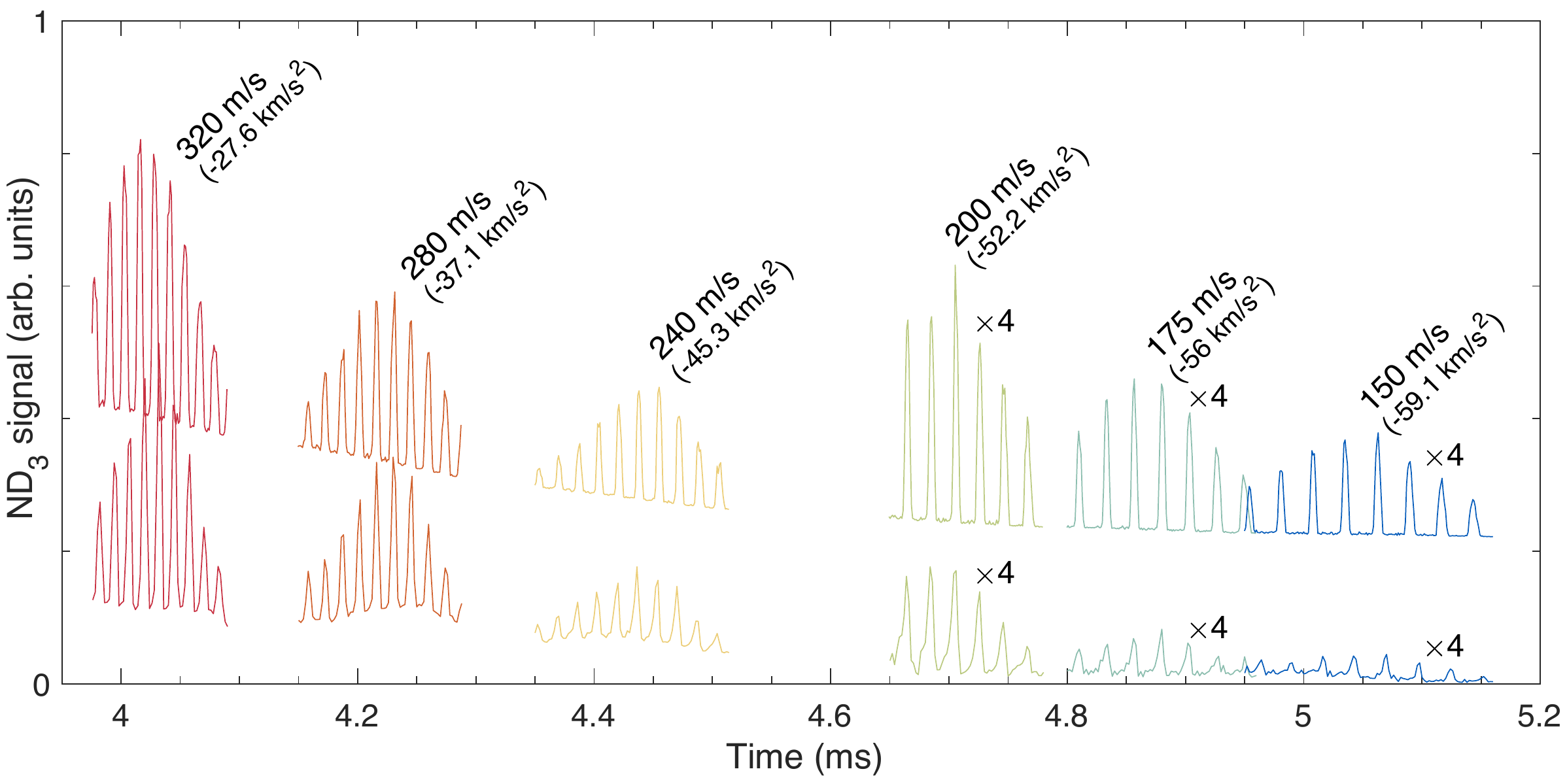}
\caption{\label{fig:prsd}Time-of-flight traces of a molecular beam decelerated with PRSD from an initial velocity of 415 m/s down to the labeled final velocities using the indicated accelerations. The three slowest packet signals have been multiplied by a factor of 4. The top row shows the results of the simulations and the bottom row shows the experimental measurements. The results of the simulations have been vertically offset for clarity.}
\end{figure*}

\begin{figure}
\includegraphics[width=3.25in]{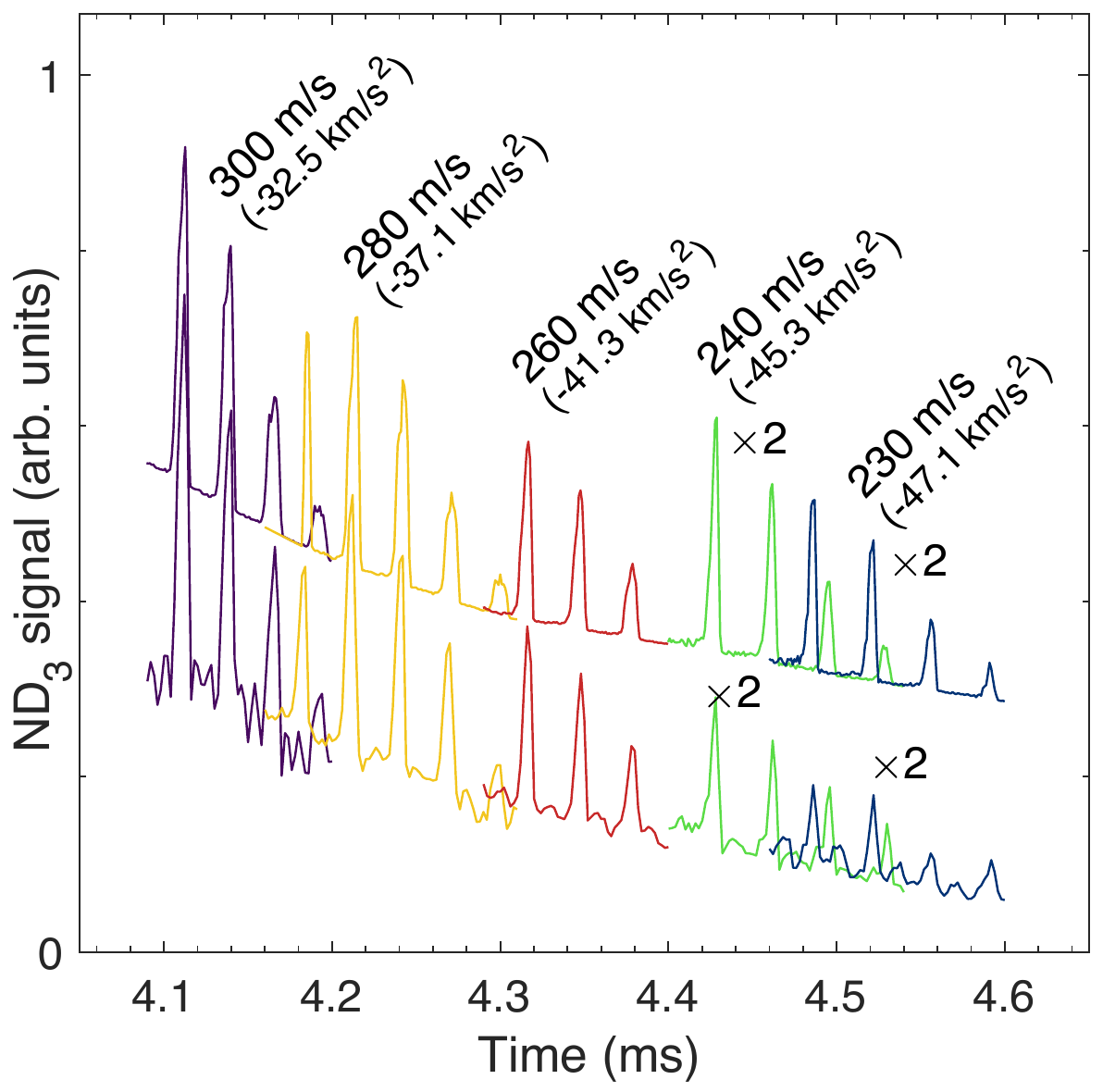}
\caption{\label{fig:twsd}Time-of-flight traces of a molecular beam decelerated with TWSD from an initial velocity of 415 m/s down to the labeled final velocities using the indicated accelerations. The two slowest packet signals have been multiplied by a factor of 2. The top row shows the results of simulations and the bottom row shows the experimental measurements. The results of the simulations have been vertically offset for clarity. }
\end{figure}

We performed experimental measurements for both PRSD and TWSD at $\pm$7 kV using the setup described in Sec. \ref{sec:expSetUp} and compared them to simulations. All experiments use a molecular beam with an initial velocity of 415 m/s. The beam was decelerated to various final velocities. Figure \ref{fig:prsd} (\ref{fig:twsd}) shows PRSD (TWSD) signals for decelerated ND$_3$ packets as a function of arrival time for selected final velocities. Both figures have results from the simulated decelerated packets on the top and the experimental data on the bottom. There are more decelerated peaks in each packet than shown, but, for clarity, only the central and a couple adjoining peaks are displayed. Both figures have the same vertical scale. The arrival time between peaks in PRSD is more closely spaced than in TWSD due to the well spanning only 2$l$ instead of 4$l$. Since there is not enough deceleration to separate the decelerated packets completely from the initial pulse, the baselines of the signals are not zero and sit on top of the low velocity tail of the initial molecular beam.  

With the same initial velocity, $v_i$=415 m/s, and moderately low accelerations decelerating to $v_f=280$ m/s, TWSD produces more decelerated ND$_3$ than PRSD. However, this is not the case at large accelerations. At $v_f<$220 m/s TWSD no longer results in observable decelerated signal, but PRSD still produces clear peaks. Although PRSD has lower signal than TWSD for $v_f>$230 m/s, it continues to produce observable decelerated molecules down to $v_f=$150 m/s. TWSD does not produce clearly decelerated packets over the same large velocity range as PRSD, and the amplitude of decelerated peaks decreases more rapidly than in PRSD as the acceleration increases. 

The simulations accurately predict the shape and arrival times of the experimental results. While the amplitude of the experimental results and simulation are comparable at higher final velocities (small acceleration), the experimental signal is lower than the simulation at lower final velocity (large acceleration). These deviations at large accelerations are likely due to imperfections in electrode rings and voltages. This may arise from individual ring misalignment or, for TWSD, deviations in amplifier output voltage from an ideal sine wave. While careful frequency tuning and calibration can mitigate amplifier fluctuations, any imperfections add jitter to the Stark potential well. At large accelerations, jitter may distort the already shallow Stark potential well to the point that many normally phase-stable molecules fall out. However, this would be less of a concern for a deeper potential well with less tilt such that imperfections in the well are small compared to the downstream Stark potential well height of the tilted well. 
\begin{figure}
\includegraphics[width=3.25in]{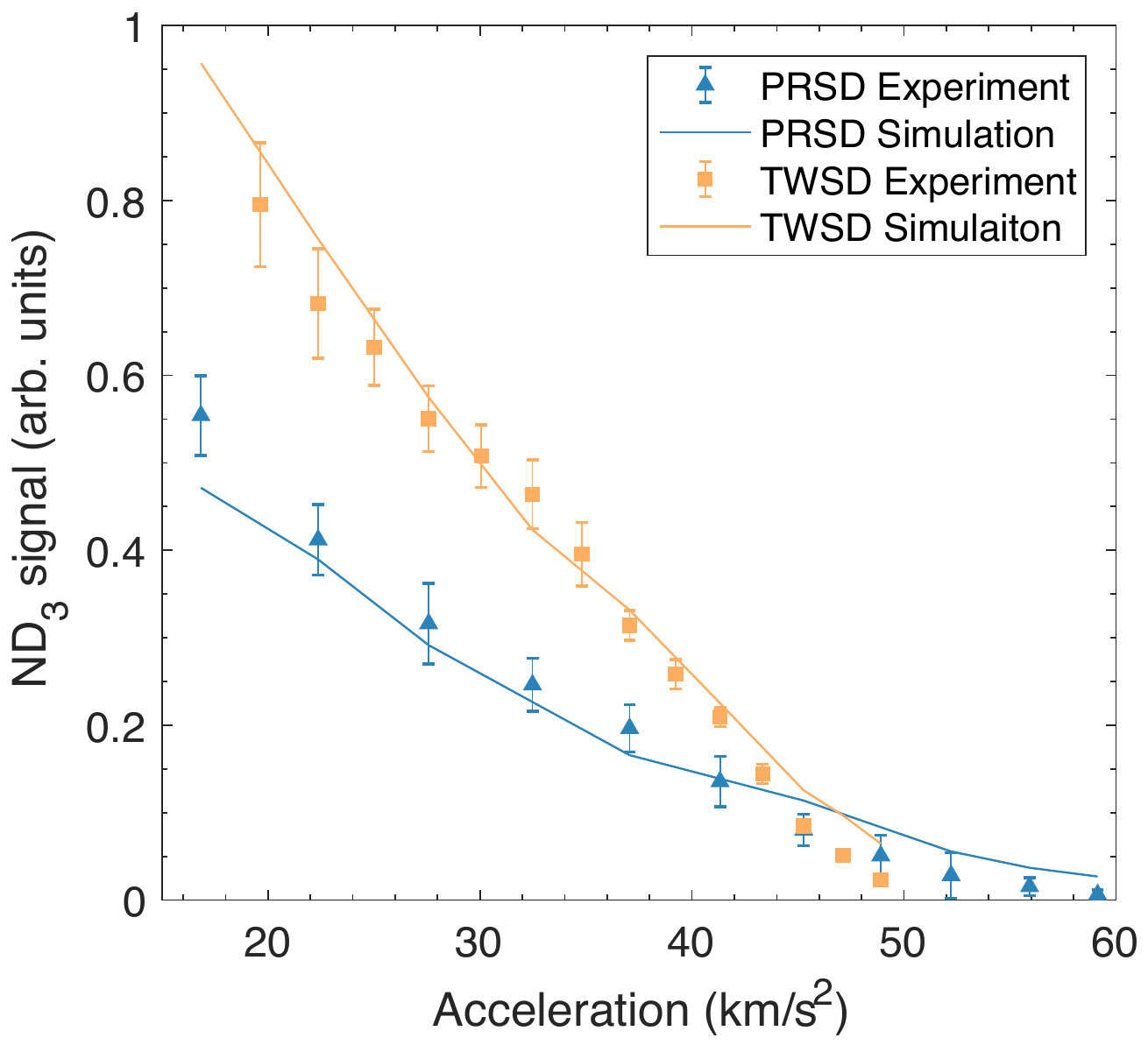}
\caption{\label{fig:7kVcross}ND$_3$ signal verses acceleration for both TWSD (yellow squares) and PRSD (blue triangles). The experimental measurements are given by individual points and the results from Monte Carlo simulations by a line. The cross over from TWSD to PRSD producing more decelerated molecule signal occurs near 46 km/s$^2$.  }
\end{figure}

We use the integrated signal above the baseline in the central peak of each packet to determine and quantify the trade-off between the two modes. The integrated signal for both experimental and simulated results is shown in Fig. \ref{fig:7kVcross}. The x-axis has been changed from final velocity to acceleration using Eq. (\ref{eq:accel}), so that the results can be applied to different initial velocities. Now, the trade-off between the two modes is clear. There is a cross-over point between the decelerated signals in TWSD and PRSD. For low accelerations TWSD yields higher overall signal, but at accelerations greater than -46 km/s$^2$ PRSD yields higher signal. Additionally, PRSD has detectable  decelerated packets at accelerations above -50 km/s$^2$ while TWSD does not. Therefore, if the desired result is a very low velocity molecular beam, PRSD is preferable for this given electrode voltage and configuration. 

\subsection{\label{subsec:phasespace}Phase-Space Acceptance}
 
\begin{figure}
\includegraphics[width=3.25in]{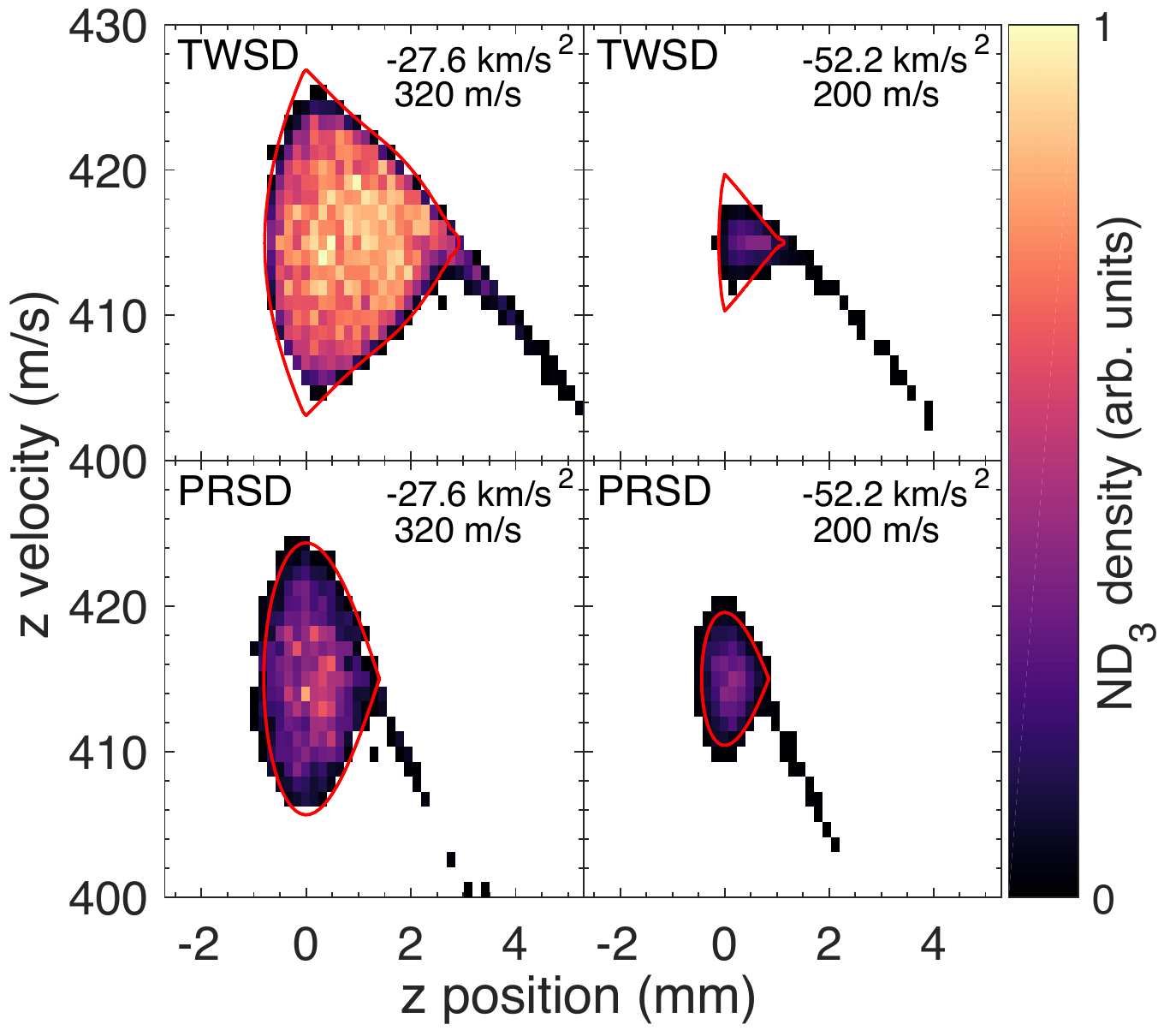}
\caption{\label{fig:phasespace}Longitudinal phase-space plots with TWSD in the top row and PRSD in the bottom row. The left column has an acceleration of -27.6 km/s$^2$ and the right column has an acceleration of -52.2 km/s$^2$. For an initial velocity of 415 m/s, this would be a final velocity of 320 m/s and 200 m/s respectively. The solid-red lines are the 1-dimensional separatricies. There is no significant structure in the phase-space acceptance in either deceleration mode. }
\end{figure}

Another method of examining the cross-over point between the two deceleration modes is to examine the phase-space acceptance, the range of initial molecule positions and velocities that will be decelerated. We define a cross-over point as the acceleration value where at higher accelerations PRSD begins to produce more decelerated signal than TWSD. One-dimensional (1D) longitudinal ($z$-axis) phase-space boundary calculations, done in the same manor as \cite{Bethlem2002}, are useful for determining the separatrix, which is the boundary between stable and unstable molecular orbits in phase space. Figure \ref{fig:phasespace} shows two-dimensional phase-space histograms of all molecules in the central decelerated peak and longitudinal phase-space separatricies for PRSD and TWSD above and below the cross-over point. The phase-space distribution has $z=0$ as the entrance of the decelerator. Only molecules from the initial distribution that are successfully decelerated and in the central peak are plotted; all other molecules are not shown.    

Both modes of deceleration have well-filled phase spaces, which differs from the structured phase spaces that are seen in PPSD \cite{vdM2006,Parazzoli2009}. Thus, the trade off in total decelerated signal comes from the change in phase-space area and not phase-space filling. From a final velocity of 320 m/s to 200 m/s the phase-space acceptance area of TWSD decreases significantly more than in PRSD. This is due to the significant decrease in Stark potential well depth from tipping due to the acceleration. While the longitudinal phase-space acceptance captures the phase-space boundary of the decelerated molecules, the area of the phase space does not accurately model the efficiency of the deceleration at different final velocities. Even a 3D phase-space volume calculation fails to model the rate of decrease in experimental signal with increased acceleration. This is because the total number of successfully decelerated molecules is determined by the overlap of the phase-space acceptance with the initial distribution of molecules. To account for the initial distribution of the beam, the deceleration efficiency should be computed using  Monte Carlo simulations.

\subsection{\label{subsec:othervolt}Determining the Optimal Deceleration Mode}
\begin{figure}
\includegraphics[width=3.25in]{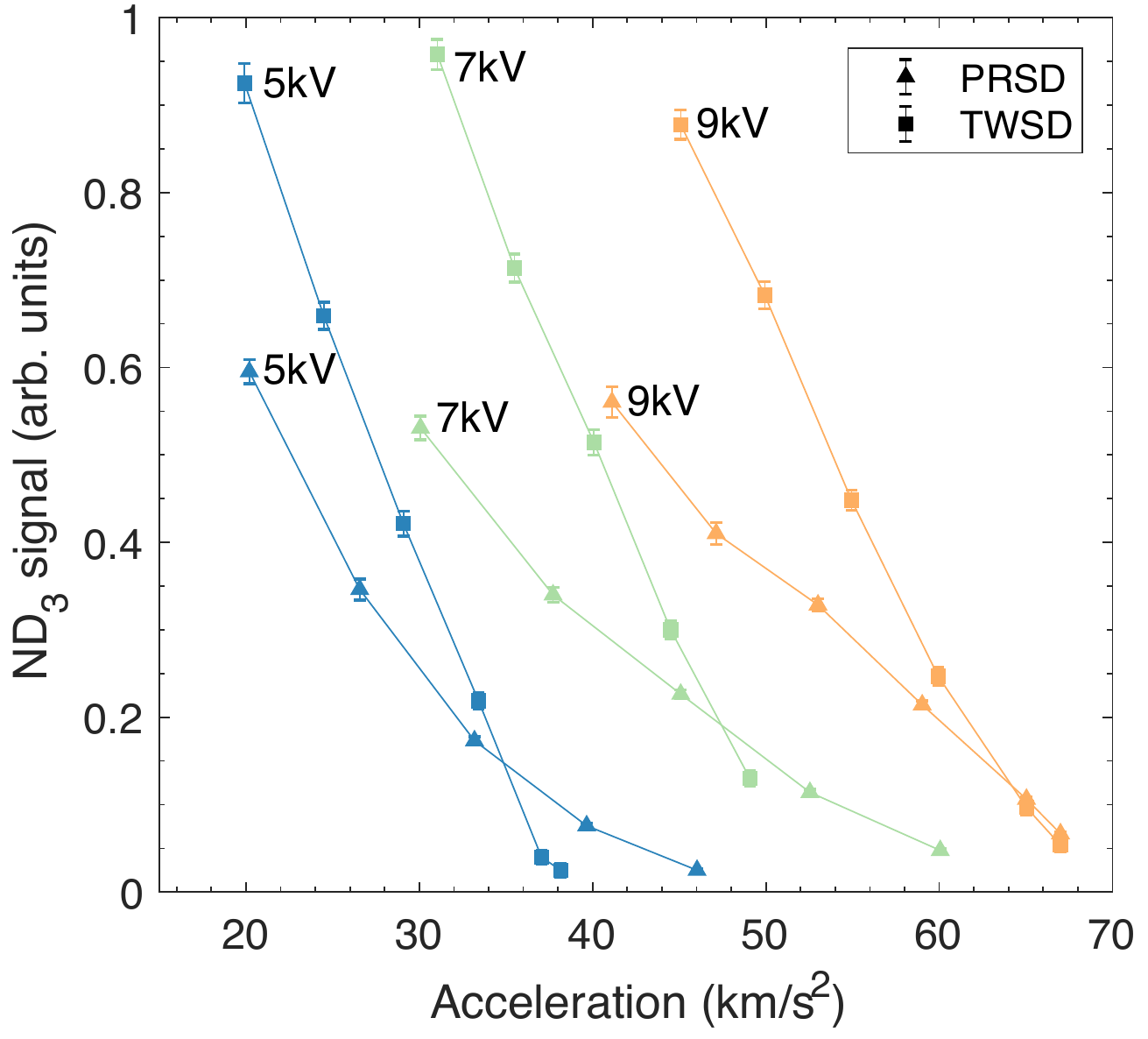}
\caption{\label{fig:crossing}Simulated integrated ND$_3$ signal verses acceleration for both TWSD (squares) and PRSD (triangles) with peak voltages of 5, 7, and 9 kV. A crossing point exists for all voltages shown.  }
\end{figure}

When using decelerated molecules in an experiment, the deceleration mode would ideally be chosen to have the maximum decelerated signal for a given final velocity. This may not always be an option due to hardware constraints. However, even if the choice is available, it is not always clear which mode is more efficient. The trade-off between the two deceleration modes is especially relevant given experimental limitations of a given decelerator. To determine which mode is best to use, different voltage regimes were explored using simulations. The same decelerator geometry as previously described was used while changing the peak voltage in both modes. Results of these studies are shown in Fig. \ref{fig:crossing} for 5, 7, and 9 kV. A crossing point where TWSD and PRSD produce the same signal exists for these voltages and occurs at a larger acceleration for greater electrode voltages. 

Generally, TWSD is more efficient for mild accelerations, but for large accelerations, especially at low peak voltages, PRSD may be more efficient. The crossing point may or may not be relevant depending on the parameters of the initial beam and maximum voltage on the electrodes. For example, while a crossing point for 10 kV (not plotted in Fig. \ref{fig:crossing}) exists, it is not accessible in our decelerator using a molecular beam seeded in krypton. The acceleration to bring ND$_3$ seeded in krypton from 415 m/s down to rest is -67.9 km/s$^2$ and the crossing point at 10 kV is near -74.3 km/s$^2$. Thus, for 10 kV on the electrodes, a beam seeded in krypton, and this decelerator geometry, TWSD will always produce more ND$_3$ at all final velocities. 

However, at lower peak voltages, PRSD is an important operating mode to consider. The advantage of the PRSD is that it allows access to higher accelerations than TWSD due to the greater Stark potential well gradient. As controlling high-voltages is one the largest experimental challenges for any Stark deceleration experiment, PRSD may be an important tool to investigate low final molecular velocities given tight voltage constraints. This is because PRSD produces more signal at higher accelerations and lower electrode voltages and is easier to implement than TWSD. PRSD also allows the beam to reach a lower final velocity, possibly one that can be trappable, without having to build a very long decelerator or obtain analog high-voltage amplifiers.

For a given decelerator length, a quick method of determining which mode will yield more decelerated molecule signal is desirable. Previously, the crossing points between the two modes were determined by integrating time-of-flight traces from Monte Carlo simulations, which is computationally intensive. Instead, the crossing point can be determined using an efficient 1D separatrix calculation.  

The 1D separatrix area does not accurately model the decrease in experimental signal with increasing acceleration for either model since it does not account for the initial molecule position and velocity spread. For the separatrix area to accurately predict the rate of decrease of the decelerated peaks the overlap of the initial distribution and the phase space must be known. However, this overlap function affects the phase space of both modes similarly and thus does not change the point where their phase-space areas are equal. Thus, calculating the acceleration when the longitudinal phase-space areas for both modes is equal gives a good estimate for where the crossing-point will occur. 

\begin{figure}
\includegraphics[width=3.25in]{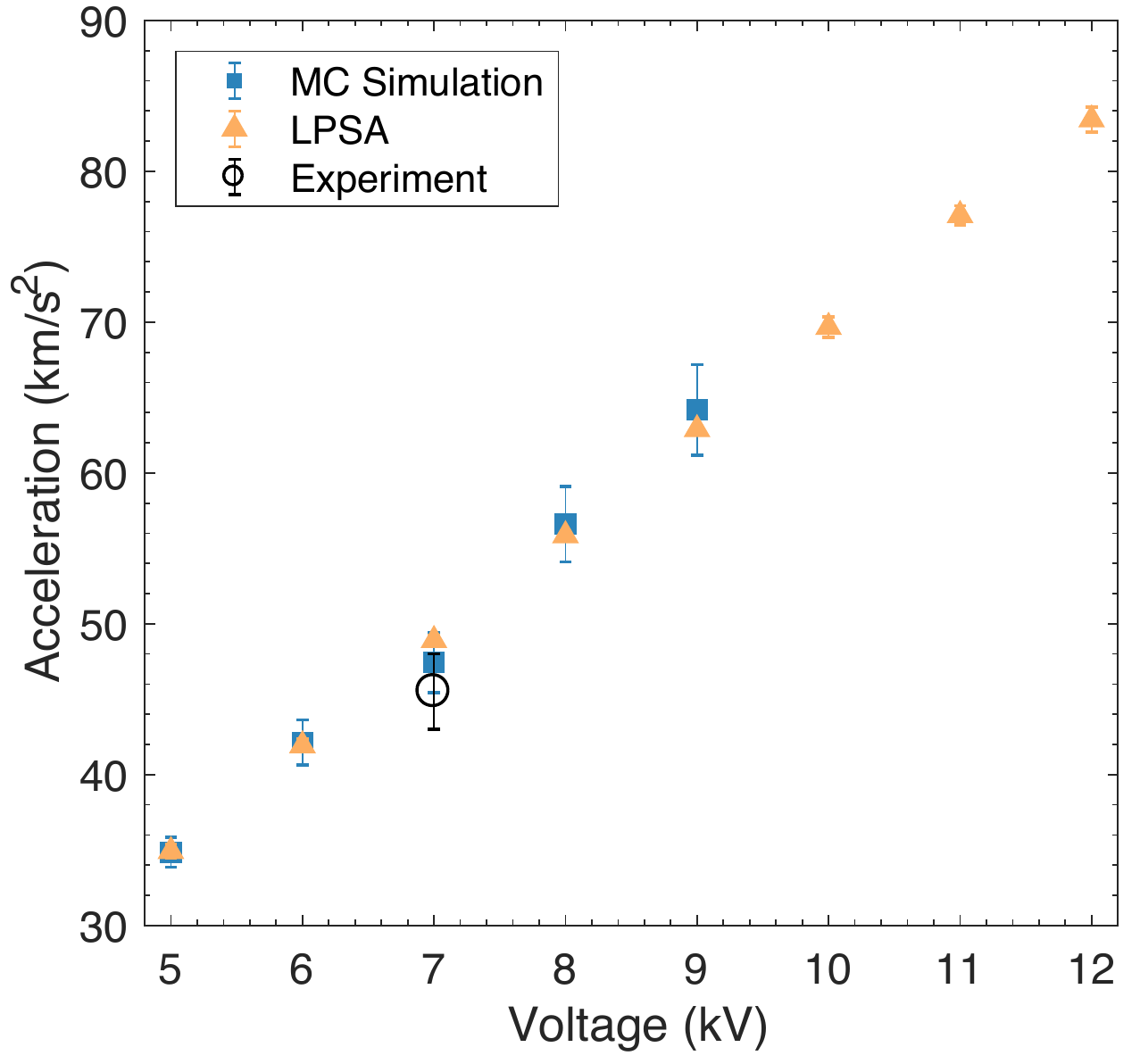}
\caption{\label{fig:linearcross} The acceleration at the crossing-point, where TWSD and PRSD produce similar decelerated molecule signal, calculated using Monte Carlo simulations and 1D longitudinal phase-space acceptance. The crossing point determined by either method is linear with respect to applied voltage. The experimental crossing-point (open circle) agrees well with the calculations and simulations.}
\end{figure}

Figure \ref{fig:linearcross} shows the crossing points determined by both Monte Carlo simulations and 1D separatrix calculations are linear with respect to applied acceleration. The experimentally determined crossing point, using the data from Fig. \ref{fig:7kVcross}, is plotted with an open circle and agrees with both calculations. The 1D separatrix calculations agree with the Monte Carlo simulations and is a simple method for determining the crossing point, even if it does not quantify how much better one mode is compared to the other. This simple 1D separatrix calculation easily predicts when one deceleration mode is more efficient than another without having to run Monte Carlo simulations.

\section{\label{sec:conclusion}Conclusion}
Stark deceleration has proved to be a reliable source of cold molecules for various applications.  However, traditional crossed-pin geometry Stark decelerators have well documented instabilities that reduce their efficiency. An alternative deceleration technology, traveling-wave Stark deceleration, is predicted to nearly eliminate the problems that reduce the number of decelerated molecules. However, the difficulty in creating the required sinusoidally varying high-voltage waveforms for TWSD has limited their integration into experiments. To possibly make implementation of a ring-geometry decelerator less challenging, we investigated running this type of decelerator in a pulsed mode, and compared it to results of running the same decelerator in a traveling-wave mode.

We demonstrated using experiential measurements and molecular trajectory simulations that for the same peak voltage on the ring electrodes, the alternative running mode, PRSD, is more efficient than TWSD at high accelerations and less efficient at low accelerations. This effect is caused by the larger Stark potential gradient for PRSD, which allows for larger accelerations before the stable phase-space region vanishes.  The crossing point at which the two modes are equally efficient depends on the maximum applied voltage, among other experimental parameters. The mode can be a particularly important consideration for lower applied voltages. Although the electronics required to run each mode have significant differences, if both deceleration modes can be implemented, each has regimes with superior performance. Depending on the desired output molecular beam parameters and available electronics, PRSD offers a potentially useful alternative method to running a ring-geometry Stark decelerator. Additionally, these results could allow for more groups to be able to implement ring-geometry Stark deceleration using the less challenging electronics, and still gain many benefits from the cylindrical symmetry of the electrodes.

\begin{acknowledgments}
We thank N. Fitch for designing and constructing the decelerator system and high-voltage amplifiers, and D. Macaluso for constructing the decelerator. This work was funded by by NSF Grant No. CHE-146997 and PHY-1734006 and AFOSR FA9550-16-0117. 
\end{acknowledgments}

\bibliography{PRvsTWpaperbib2}
\end{document}